\documentclass[final,3p,times,twocolumn]{elsarticle}
\usepackage{graphicx}
\usepackage{amssymb}
\biboptions{sort&compress}
\usepackage{algorithmic}
\usepackage[ruled,vlined,commentsnumbered]{algorithm2e}
\usepackage{multirow}
\usepackage{array}
\usepackage{mdwmath}
\usepackage{mdwtab}
\usepackage{eqparbox}
\usepackage{url}
\usepackage[cmex10]{amsmath}

\journal{Artificial Intelligence in Medicine}

\begin{document}

\begin{frontmatter}


\title{An Attention-based Weakly Supervised framework for Spitzoid Melanocytic Lesion Diagnosis in WSI}

\ead{madeam2@upvnet.upv.es}



\author{Roc\'io del Amor\textsuperscript{1}, Laëtitia Launet\textsuperscript{1}, Adri\'an Colomer\textsuperscript{1}, Anaïs Moscardó\textsuperscript{2}, Andrés Mosquera-Zamudio\textsuperscript{2}, Carlos Monteagudo\textsuperscript{2} and Valery Naranjo\textsuperscript{1}}

\address{\textsuperscript{1}Instituto de Investigación e Innovación en Bioingeniería, Universitat Politècnica de València, 46022, Valencia, Spain}
\address{\textsuperscript{2}Pathology Department. Hospital Clínico Universitario de Valencia, Universidad de Valencia, Valencia, Spain}

\begin{abstract}

Melanoma is an aggressive neoplasm responsible for the majority of deaths from skin cancer. Specifically, spitzoid melanocytic tumors are one of the most challenging melanocytic lesions due to their ambiguous morphological features. The gold standard for its diagnosis and prognosis is the analysis of skin biopsies. In this process, dermatopathologists visualize skin histology slides under a microscope, in a high time-consuming and subjective task. In the last years, computer-aided diagnosis (CAD) systems have emerged as a promising tool that could support pathologists in daily clinical practice. Nevertheless, no automatic CAD systems have yet been proposed for the analysis of spitzoid lesions. Regarding common melanoma, no proposed system allows both the selection of the tumoral region and the prediction of the diagnosis as benign or malignant. Motivated by this, we propose a novel end-to-end weakly-supervised deep learning model, based on inductive transfer learning with an improved convolutional neural network (CNN) to refine the embedding features of the latent space. The framework is composed of a source model in charge of finding the tumor patch-level patterns, and a target model focuses on the specific diagnosis of a biopsy. The latter retrains the backbone of the source model through a multiple instance learning workflow to obtain the biopsy-level scoring. To evaluate the performance of the proposed methods, we perform extensive experiments on a private skin database with spitzoid lesions. Test results reach an accuracy of 0.9231 and 0.80 for the source and the target models, respectively. Besides, the heat map findings are directly in line with the clinicians’ medical decision and even highlight, in some cases, patterns of interest that were overlooked by the pathologist due to the huge workload.

\end{abstract}

\begin{keyword}
Spitzoid lesions \sep Attention convolutional neural network \sep Inductive transfer learning \sep Multiple instance learning \sep Histopathological whole-slide images 
\end{keyword}

\end{frontmatter}

%\linenumbers

%% SECTIONS
\section{Introduction}\label{intro}

According to the World Health Organization, nearly one in three diagnosed cancers is a skin cancer \cite{Apalla2017}. Although two of the most diagnosed skin cancers are basal cell carcinoma and squamous cell carcinoma, non-melanocytic lesions developed from non-pigmented skin cells, the most dangerous skin cancer is melanoma which is responsible for 80 percent of skin cancer-related deaths \cite{articulo_carlos}.

%with a mortality rate of 1.62\% among other skin cancers \cite{li2020skin}. 

Melanoma is an aggressive melanocytic neoplasm with numerous resistance mechanisms against therapeutic agents. In most melanocytic tumors, a precise pathological distinction between benign (nevus) and malignant (melanoma) is possible. However, there are still uncommon melanocytic lesions that represent a diagnostic challenge for pathologists because of their ambiguous morphological features. Among these, one of the most challenging lesions is the so-called `spitzoid melanocytic tumors' (SMTs), composed of spindled/epithelioid melanocytes with a large nucleus \cite{Wiesner2016}. SMTs are particularly difficult to diagnose in regard to malignancy, partly because of the discrepancy between the histopathological morphology and the clinical evolution. %As a consequence of the conflicting morphological features, the terms “atypical Spitz tumors” (ASTs) or “spitzoid tumor of uncertain malignant potential" (STUMP) encompass spitzoid melanocytic lesions that show overlapping characteristics with both Spitz nevus and spitzoid melanoma \cite{Wiesner2016}. The resulting ambiguity of UMP malignancy may lead to uncertainty and error in diagnosis and therefore may hinder appropriate clinical treatment \cite{Menezes2017}. The STUMP prognosis represents a formidable diagnostic challenge for dermatopathologists and therefore, it is the hottest open area in SMT detection.

\begin{figure*}[h]
    \begin{center}
    \includegraphics[width=\textwidth]{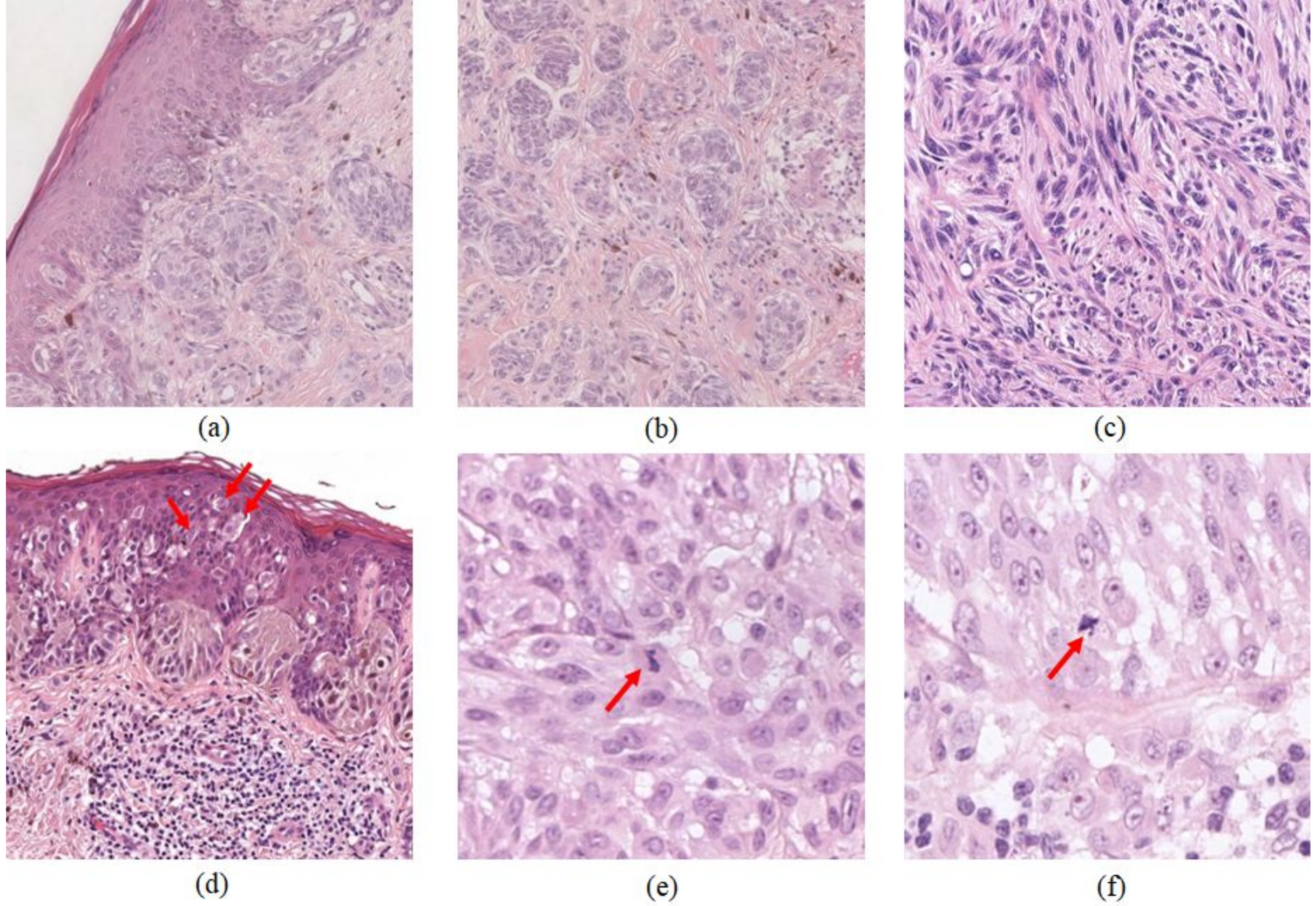}
    \end{center}
    \caption{Representative patches extracted from WSIs presenting different spitzoid melanocytic lesions; (a)-(b): Benign spitzoid nevus containing melanocytic nests in an organized fashion; (c): Malignant lesion representative of the cellular disorder; (d): Malignant tissue with pagetoid spread, very common in this type of lesions; (e)-(f): Typical and atypical mitoses.}
    \label{fig:patchesLesions}
\end{figure*}

The final diagnosis of SMTs, as for most melanocytic and non-melanocytic lesions, is confirmed by skin biopsies. The skin tumor is excised, laminated, stained with Hematoxylin and Eosin (H\&E) and finally stored in crystal slides. Then, dermatopathologists analyze the sample under the microscope \cite{Wiesner2016}. During the analysis of spitzoid lesions, different histopathological characteristics can be observed depending on the malignancy degree, see Figure \ref{fig:patchesLesions}. The regions with spitzoid nevus lesions generally have a confluence of organized melanocyte nests. Figure \ref{fig:patchesLesions} (a)-(b) shows sub-regions of a benign spitzoid melanocytic lesion. These regions show organized and well-grouped melanocytes with cell maturation (melanocytes decrease in size towards the base of the lesion) distributed throughout the dermal area. In this case, this type of nevus lesion is known as compound nevus. If the lesion only occurs in the epidermis and does not show extension into the dermis it would be called junctional nevus. In the case of spitzoid melanoma lesions, the nests are no longer organized as in benign lesions, see Figure \ref{fig:patchesLesions} (c). In this region, we cannot distinguish melanocyte nests as well defined as in Figure \ref{fig:patchesLesions} (a)-(b) (benign lesions). Therefore, in malignant lesions, the cellular disorder is a frequent pattern. More features associated with malignancy of spitzoid melanocytic lesions include cytological atypia with marked nuclear pleomorphism, pagetoid spread (melanocytic cells grow and invade the upper epidermis from below) and a poor circumscription of lesions at their peripheries \cite{Barnhill2006}. Figure \ref{fig:patchesLesions} (d) shows an example of the pagetoid pattern located in the epidermis. In addition to the cellular disorder and the presence of small melanocytic nests, there are other local-level features associated with signs of malignancy. Among these malignant patterns stand out typical and atypical mitoses. Note that benign melanocytic lesions can also have typical mitoses, particularly in the most superficial area. Therefore, if we find a typical mitosis in a spitzoid lesion, we should take into account more factors such as the amount of mitoses and their location in the lesion (mitoses in the depths of the lesion have a worse prognosis) to determine if the neoplasm is malignant. Typical mitoses are thus a sign of malignancy but are not determinant in establishing that a neoplasm is malignant. However, if we find an atypical mitosis in a spitzoid lesion, it is almost certain that it will be diagnosed as malignant. An example of typical and atypical mitoses on a malignant lesion are shown in Figure \ref{fig:patchesLesions} (e)-(f), respectively. The presence of the aforementioned morphological characteristics allows pathologists to diagnose histological images under the microscope. However, this process is highly time-consuming and can lead to discordance between histopathologists due to the ambiguity of these neoplasms \cite{Lodha2008}. This is the reason why these lesions represent a formidable diagnostic challenge.

The computer-aided diagnosis systems (CADs) aim to support pathologists in the daily analysis of skin biopsies, reducing both the workload and the inconsistency generated. With the emergence of digital pathology, the digitization of histological crystals into whole-slide images (WSIs) has been standardized \cite{Gurcan2009}, leading the way to the application of computer vision methods. The development of CADs based on WSI analysis presents important hardware limitations because of their large size. This is the reason why the typical approach generally involves extracting small patches from larger WSIs, resulting in thousands of patches per image. Previous research successfully showed the applicability of computer vision methods based on deep-learning approaches by means of convolutional neural networks (CNN) to WSI interpretation following a patch-based strategy. The CNN-based approaches have been extensively tested for the detection of breast cancer \cite{Bejnordi2017, Rakhlin2018, Litjens2016, Wang2016}, prostate cancer \cite{Silva-Rodriguez2020, JimenezdelToro2017, Litjens2016} or lung cancer \cite{Yu2016, Coudray2018}, using WSIs. However, regarding skin cancer diagnosis, specifically for melanoma detection, most research was based on the analysis of dermoscopic images \cite{codella2017deep,Esteva2017,haenssle2018man, Maron2019,brinker2019deep,kassani2019comparative, Liu2020,astorino2020melanoma,yu2018acral} and few studies have focused on the analysis of WSIs  \cite{Hekler2019, DeLogu2020,wang2020automated,Belaala2020}. In Hekler et al. \cite{Hekler2019}, patches were obtained from randomly cropped WSIs. Specifically, one crop per slide/patient was obtained. Transfer learning was used on a pre-trained ResNet50 CNN to differentiate between two classes, benign and melanoma tissues. This study aimed to prove that the developed model showed pathologist-level discordance despite using relatively few data. However, the main limitation of this work is that they are not able to analyze entire WSIs but only a characteristic tumor sub-region. In De Logu et al. \cite{DeLogu2020}, more than 3000 patches were extracted from regions of interest (ROIs) of 100 WSIs. A pre-trained Inception-ResNet-v2 network was then used to distinguish cutaneous melanoma areas from healthy tissues, showing potential to help pathologists and improve tumor identification accuracy. However, this work didn't discriminate melanoma from nevi WSIs. In \cite{wang2020automated}, the authors developed a deep-learning system to automatically detect malignant melanoma in the eyelid from histopathological sections. The authors used the VGG16 model to assign patch-level classification. Using the malignant probability from the CNN, the patches were embedded back into each WSI to generate a visualization heatmap and leveraged a random forest model to establish a WSI-level diagnosis between malignant and benign. The main limitation of this work is that the input of the algorithm is the tumoral region and not the entire WSI image. Specifically, for spitzoid melanocytic lesions, since they represent an uncommon group of tumors, very few studies have applied computer vision techniques to their analysis yet. 

To the best of the authors' knowledge, no previous studies were focused on the SMTs distinction based on data-driven approaches. There is only one method based on hand-crafted feature extraction for SMTs identification \cite{Belaala2020}. In \cite{Belaala2020}, the authors used a machine-learning algorithm to assist in the diagnosis of SMT. In this study, a random forest classifier was used on numerical morphological characteristics extracted by the pathologists from histological images \cite{Belaala2020}. Therefore, the method does not extract features directly from the histological images. %This approach only considered the distinction between Spitz nevus and UMP, leaving aside spitzoid melanoma classification. Additionally, this works does not propose an eventual prognosis for ASTs' malignant potential in malignant o nevus neoplasm. 
As SMTs are uncommon skin lesions, the available data is generally scarce. This is the reason why this study used data from 54 patients. %, of which 45 corresponded to Spitz nevus and 7 to UMP.

Inspired by the main limitations of the studies focused on the melanoma detection and more specifically on SMTs diagnosis, in this work, we put forward a novel semi-supervised inductive transfer learning strategy to conduct both the local automatic detection of tumor regions and the global prediction of an entire biopsy. To the best of our knowledge, this is the first attempt to detect spitzoid melanocytic lesions within WSIs in an automatic way. In summary, the main contributions of this work are:

\begin{itemize}

   \item Spitzoid histological images are used for the first
    time to develop an automatic feature extractor.
    
    \item A new histology-based backbone is proposed to extract more accurate features.
   
    \item A novel framework based on inductive transfer learning to solve at the same time ROI selection and malignancy detection is developed.
    
     \item Multiple instance learning-based solutions are formulated in a novel framework for spitzoid lesion detection.

     \item A wide clinical interpretability of the results achieved with the proposed methods is provided.
     
\end{itemize}

The rest of the paper is structured as follows. Section 2 details the related work regarding inductive transfer learning and multiple instance learning strategies, the underlying methodologies of the present work, to go on to underline the improvement introduced in the medical research. In Section 3, we present the data used in this work, CLARIFYv1, a private database comprised of skin WSIs from patients with spitzoid tumors. In Section 4, we describe the proposed methodology, mainly composed of two stages: i) development of a source model in charge of performing a patch-level classification to select tumor regions and ii) a target model based on a multiple instance learning approach to predict the malignancy degree at the biopsy level. Sections 5, 6 and 7 provide information on the performance outcomes related to the different classification tasks. Finally, in Section 8 we present our conclusions along with the future work.

\section{Related work}

\begin{enumerate} [A.]
\item \textit{Inductive transfer learning}
\end{enumerate}

Given a source domain $D_S$ with a corresponding source task $T_S$, and a target domain $D_T$ with a corresponding task $T_T$, transfer learning (TL) is the process of improving the target predictive function $f_T(\cdot)$ by using the related information from $D_S$ and $T_S$, where $D_S \neq D_T$ or $T_S\neq T_T$ \cite{weiss2016survey}. In the context of this work, we refer to inductive transfer learning (ITL) as the ability of the learning mechanism to enhance the performance on the target task (with a reduced number of labels) after having learned a different but related concept or skill on a previous task in the same domain \cite{Vilalta2010}. The intuition behind this idea is that learning a new task from related tasks should be easier, faster and with better solutions or using less amount of labeled data than learning the target task in isolation. When the source and the target domain labels are available, the inductive transfer learning approach is known as multi-task learning.

Interest in this technique has grown in recent years in applications related to medical issues due to the promising results obtained. In this context, Caruana et al. suggested using multi-task learning in artificial neural networks and proposed an inductive transfer learning approach for pneumonia risk prediction \cite{caruana1997multitask}. Silver et al. introduced a task rehearsal method (TRM) as an approach to life-long learning that used the representation of previously learned tasks as a source of inductive bias. This inductive bias enabled TRM to generate more accurate hypotheses for new tasks that have small sets of training examples \cite{silver2002task}. Zhang et al. used a technique based on inductive transfer learning to solve two-step classification problems: classification of malignant-nodule and non-nodule, and to classify the Serious-Malignant and the Mild-Malignant in malignant-nodule \cite{zhang2019computer}. Tokuoka et al. provided an inductive transfer learning approach to adopt the annotation label of the source domain datasets to tasks of the target domain using Cycle-GAN based on unsupervised domain adaptation (UDA) \cite{tokuoka2019inductive}. Zhou et al. used an inductive transfer learning method to improve the performance of ocular multi-disease identification. In this case, the source and the target domain data were fundus images, but the source and target domain tasks were diabetic retinopathy lesion segmentation and multi-disease classification, respectively \cite{zhou2020benchmark}. De Bois et al. used an inductive transfer learning approach to build a better glucose predictive model using a CNN-based architecture. A first model was trained on source patients that may come from different datasets and then, the model was fine-tuned to the target patients. Adding a gradient reversal layer, the patient classifier module made the feature extractor learn a feature representation that was general across the source patients \cite{de2021adversarial}.

In that context, we adopt an inductive transfer strategy to accurately classify instances from WSIs. The source model is trained to predict tumor regions by a patch-based CNN using inaccurate annotations with a large number of labels. After that, the backbone of the source model is retrained to classify nevus and malignant biopsies using a target model where the number of labels is reduced as this model is retrained at the biopsy level.

\begin{enumerate} [B.]
\item \textit{Multiple instance learning}
\end{enumerate}

Multiple instance learning (MIL), a particular form of weakly-supervised learning, aims at training a model using a set of weakly labeled data \cite{srinidhi2020deep}. In MIL tasks, the training dataset is composed of bags, where each bag contains a set of instances. A positive label is assigned to a bag if it contains at least one positive instance. The goal of MIL is to teach a model to predict the bag label. MIL approach has been successfully applied to computational histopathology for tasks such as tumor detection based on WSIs, reducing the time required to perform precise annotations \cite{campanella2019clinical,das2020detection,zhao2020predicting, silva2021self}.
%yao2019deep, hashimoto2020multi
In this vein, \cite{campanella2019clinical,das2020detection} assigned the global label (cancerous against non-cancerous) to all patches of a slide. Campanella et al. \cite{campanella2019clinical} proposed a MIL-based deep learning system to accomplish the identification of three different cancers: prostate cancer, basal cell carcinoma and breast cancer metastases. In this case, they used an instance-level paradigm obtaining a tile-level feature representation through a CNN. These representations were then used in a recurrent neural network to integrate the information across the whole slide and report the final classification result to obtain a final slide-level diagnosis. Das et al. \cite{das2020detection} used an embedded-space paradigm based on multiple instance learning to predict breast cancer. Specifically, they used a deep CNN architecture based on the pre-trained VGG-19 network to extract the features of each bag. Then, the bag level representation is achieved by the aggregation of the features through the batch global max-pooling (BGMP) layer at the feature embedding dimension. Silva et al \cite{silva2021self} used a novel weakly-supervised deep-learning model, based on self-learning CNNs, that leveraged only the global Gleason score of gigapixel whole slide images during training to accurately perform both, grading of patch-level patterns and biopsy-level scoring. Other works like \cite{zhao2020predicting} treated the tumor areas manually annotated by pathologists as a bag. In this case, the authors proposed a MIL method based on a deep graph convolutional network and feature selection for the prediction of lymph node metastasis using histopathological images of colorectal cancer. To the best of the authors' knowledge, no previous works have taken advantage of the promising MIL-based approaches for the diagnosis of melanocytic tumors yet. Our starting surmise is that since there is at least one identifying patch of malignancy in a melanoma lesion, the MIL-based approach could assist in diagnosing a spitzoid lesion based on its whole context lessening the ambiguity between malignant and benign lesions. Additionally, in contrast to the works cited above, as in this study each bag contains the tumoral region pseudo-labeled by the source model, the number of noisy labels is reduced, which will facilitate model training-loop since the number of available samples is particularly limited.

\section{Materials}
\label{material}

To evaluate the proposed learning methodology, we resort to a private database, CLARIFYv1, containing histopathological skin images from different body parts with spitzoid melanocytic lesions. %After analyzing the literature, no skin spitzoid tissue image databases were found. Thus, the database presented in this paper will be publicy available after the publication of this paper for further research.
The database is composed of 53 biopsies from 51 different patients who signed the pertinent informed consent. The number of patients used in this study is relatively limited because these lesions are uncommon among the population. The tissue samples were sliced, stained and digitized using the Ventana iScan Coreo scanner at 40x magnification obtaining WSIs. The slides were analyzed by an expert dermatopathologist at the Hospital Clínico of Valencia. Specifically, 21 of the 51 patients under study were diagnosed as malignant melanocytic lesions (melanoma) and the rest as benign melanocytic lesions (nevus). 

\begin{figure*}[htb]
    \centering
    \includegraphics[width=\textwidth]{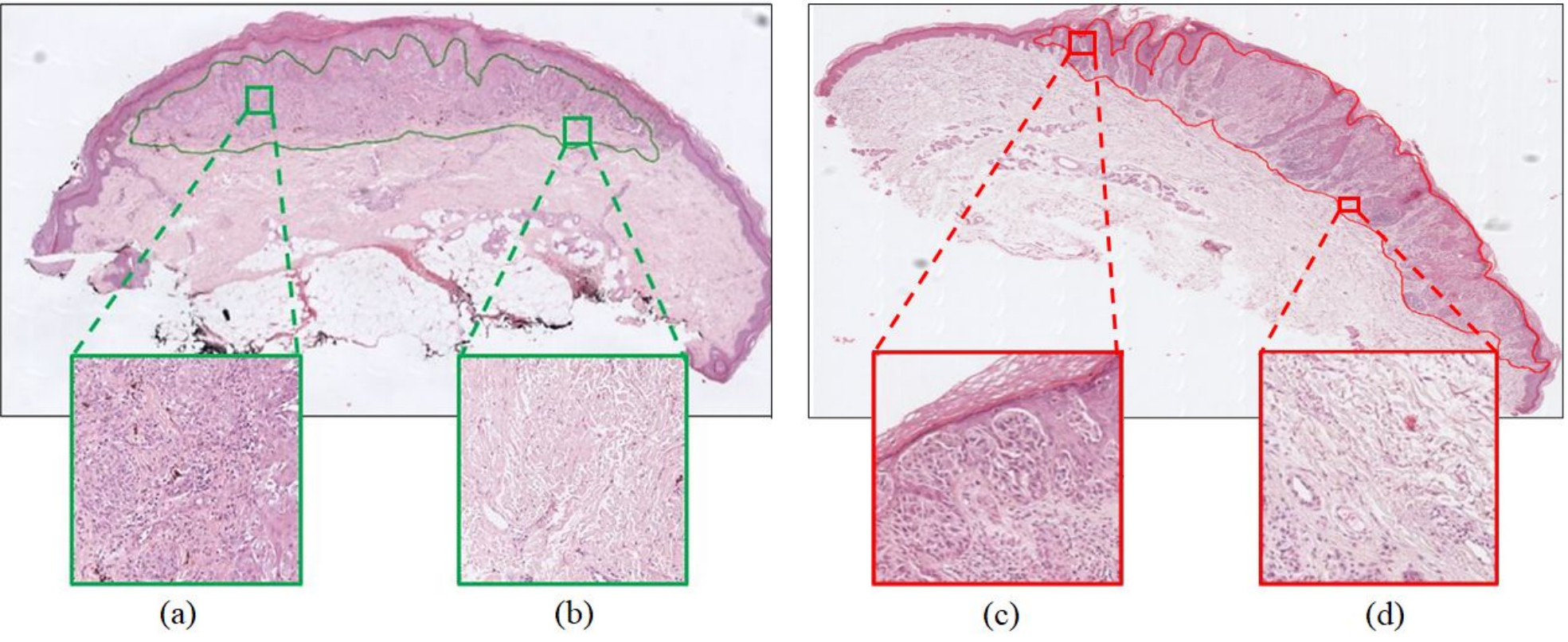}
    \caption{Annotation of a benign and a malignant tissue. Patches (a) and (c) show characteristic patterns of the tumor region, benign and malignant respectively. Although patches (b) and (d) belong to the interest region annotated by the pathologists, these patches are not characteristic of the tumor and resemble healthy tissue.}
    \label{fig:anotation}
\end{figure*}

The global tumor regions were annotated by the pathologist using an in-house software based on the OpenSeadragon libraries \cite{Openseadragon}. Note that with these annotations, WSIs were divided into interest regions (tumor region) and non-interest regions (the rest of the WSI). Figure \ref{fig:anotation} shows the annotation of benign and malignant regions. To streamline the annotation task, these annotations were performed in a coarse way, so in some sub-regions there are tumor discontinuities not considered. This fact is shown in Figure \ref{fig:anotation} (b) and (d), where these patches have not patterned related to the tumor lesion. 

In order to process the large WSIs, these were downsampled to 10$x$ resolution, divided into patches of size 512x512x3 with a 50\% overlap among them. Aiming at pre-processing the biopsies and reduce the noisy patches, a mask indicating the presence of tissue in the patches was obtained by applying the Otsu threshold method over the magenta channel. Subsequently, the patches with less than 20\% of tissue were excluded from the database. A summary of the database description is presented in Table \ref{tab:referece_patches}. 
Note that, due to the irregular morphology of these lesions, the tumor shape is very different among patients varying considerably the number of patches per patient. 

\begin{table}[hbt]
\caption{CLARIFYv1 database description. Amount of whole slide images with their respective biopsy label (first row), number of patches of each tumor region (second row) and number of non-interest region (third row).}
\label{tab:referece_patches}
\centering
\resizebox{7.6cm}{!}{%
\begin{tabular}{lcc}
\hline
                                                                           & \multicolumn{1}{l}{\textbf{Benign}} & \multicolumn{1}{l}{\textbf{Malignant}}  \\ \hline
\textbf{\# WSI}                                                            & 30                                  & 21                                                                   \\
\textbf{\begin{tabular}[c]{@{}l@{}}\# Tumor patches \end{tabular}} &  3652                                   &     4726                                                                   \\
\textbf{\begin{tabular}[c]{@{}l@{}}\#  Non tumor patches \end{tabular}}   &    5842                                 &    8139                                                                      \\ \hline
\end{tabular}}
\end{table}
\section{Methodology}

\begin{figure*}[htb]
\centering
\includegraphics[width=\textwidth]{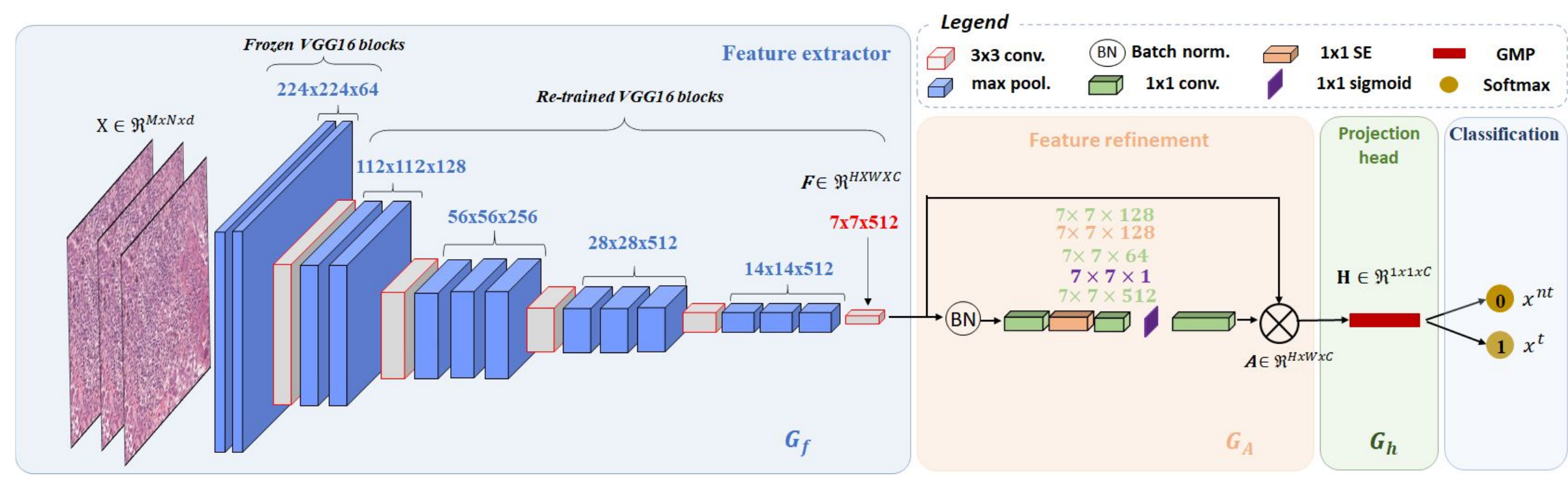}
\caption{Overview of the proposed source model to conduct the tumor region detection. Blue and orange frames correspond to the base encoder network consisting of feature extraction and refinement. Note that VGG16 has been used as the feature extractor. After that, a projection head (green frame) maps the embedded representations in a lower-dimensional space to maximize the agreement in the classification stage (cyan frame). }
\label{Source_model}
\end{figure*}

\begin{figure*}[htb]
\centering
\includegraphics[width=\textwidth]{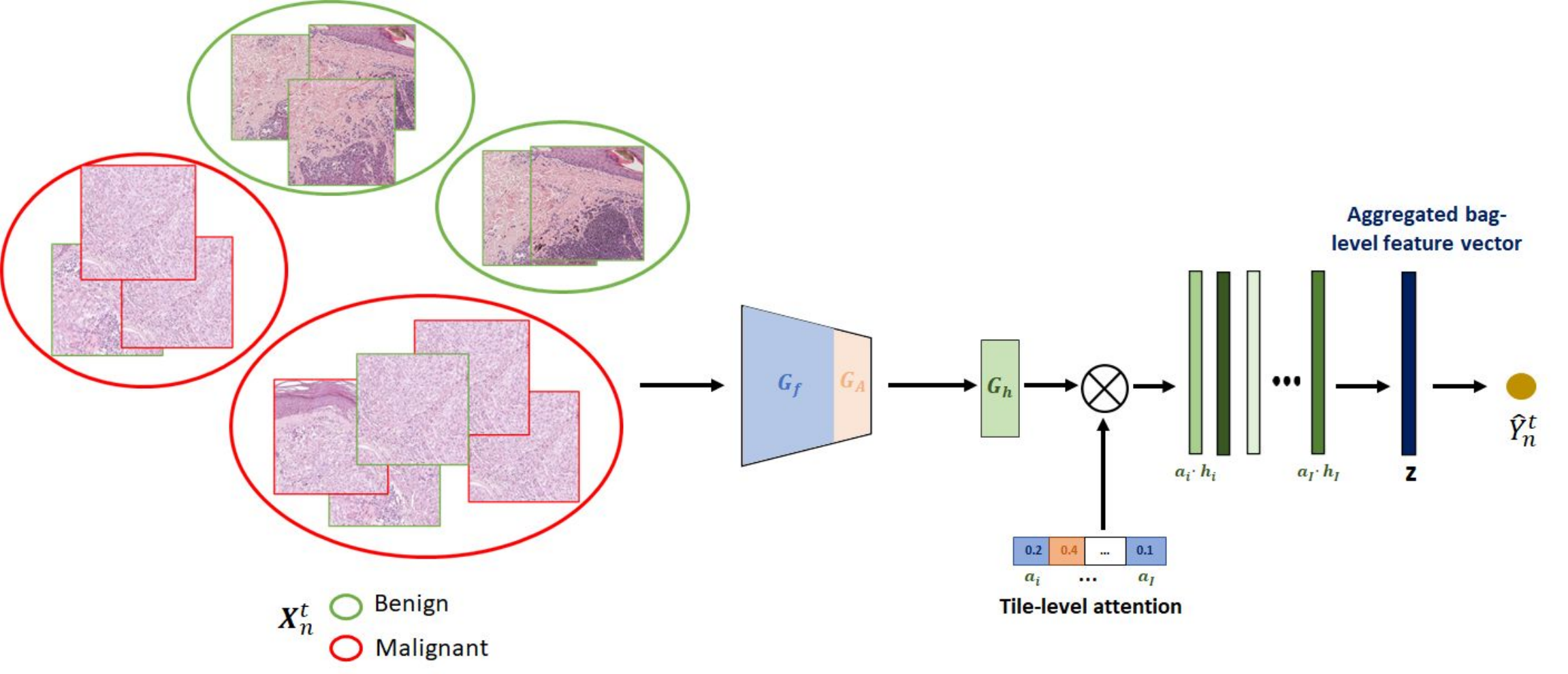}
\caption{ Pipeline showing the embedded-level approach for spitzoid melanocytic lesion classification. The weights of the pre-trained feature extractor and feature refinement of the source model ($\sigma^s$ and $\delta^s$) are used to initialize this approach. After that, we use the output of the projection head and tile-level attention to weight the patches in the prediction of a whole biopsy. Using an aggregated bag-level feature vector we classify the entire biopsy. }
\label{target_model}
\end{figure*}

The methodological core of the proposed approach is a semi-supervised CNN classifier able to detect the tumor region in a WSI and classify it into either benign or malignant spitzoid lesions. The proposed workflow is composed of a source and a target model,  ($\theta^s$) and ($\theta^t$) respectively. The first model ($\theta^s$) allows to automatically obtain the patches with significant features of spitzoid neoplasms, Figure \ref{Source_model}. Tumor patches selected by the first model are then transferred to a second model ($\theta^t$), Figure \ref{target_model}. This second model discerns malignant and benign biopsies using a MIL paradigm. 

\subsection{Source model: ROI selection}
\label{feature_extractor}

The objective of this stage is to build a 2D-CNN architecture able to extract discriminatory features from WSI patches to distinguish tumor regions. 

\begin{enumerate} [A.]
\item \textit{Backbone}
\end{enumerate}

\textit{(1) Feature extractor}. The patch-level feature extractor $G_f:x \rightarrow F$ is a CNN which maps an image \textit{x} into an \textit{F} feature volume. Since the deep-learning models trained from scratch report worse performance in comparison to fine-tuned models when the amount of available data is limited, we fine-tuned several well-known architectures: VGG16 \cite{simonyan2014very}, ResNet-50 \cite{he2016deep}, InceptionV3 \cite{szegedy2016rethinking} and MobileNetV2 \cite{huang2017densely}. All of them were pre-trained with around 14 million natural images corresponding to the ImageNet dataset. For the feature extraction stage, the base model is extracted from those pre-trained models and partially retrained. Since the patterns of the ImageNet dataset are very different from the histological ones, it is optimal to keep the low-level features only (contours, combination of basic colors, general shapes, etc.). To this end, the weights of the first convolutional blocks from the pre-trained model are frozen, while the rest are re-trained to adapt the model to the specific application. The layer from which the freezing strategy is applied is empirically optimized for each architecture and it is specified in the experimental part of the paper, Section 5. Therefore, given a histological image $x\in \mathbb{R}^{M \times N \times d}$, where $M \times N \times d = 224 \times 224 \times 3$, a feature embedded map $F\in \mathbb{R}^{H \times W \times C}$ is provided by the feature extractor. It is denoted as $ F=G_f(x;\sigma^s)$ where $\sigma^s$ is the set of trainable parameters of this source model.

\begin{figure}[h]
\centering
\includegraphics[height=1.8cm]{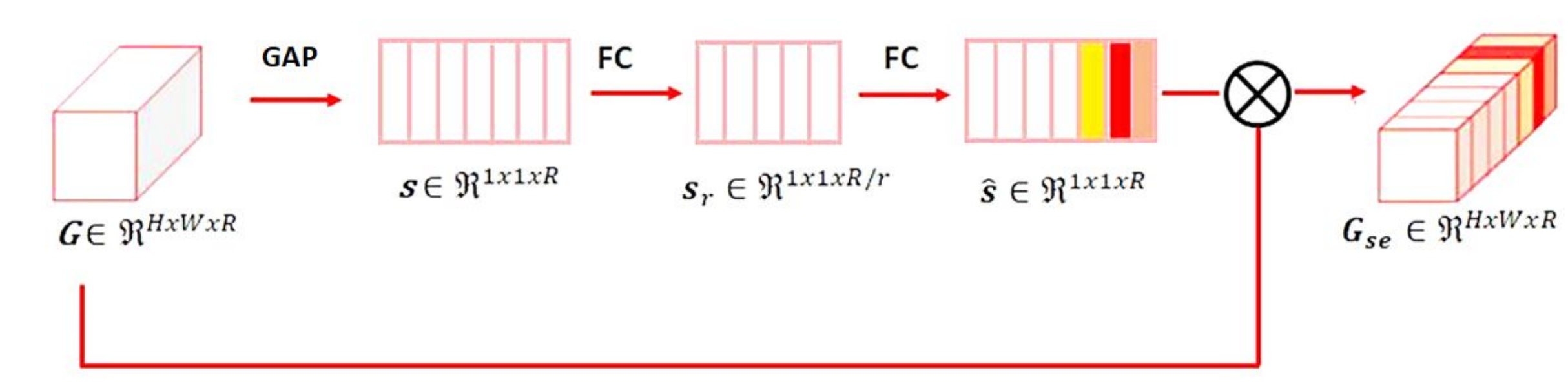}
\caption{Architecture of the Squeeze-and-Excitation blocks used to exploit the dependencies between feature channels.}
\label{SE}
\end{figure}

\textit{(2) Feature refinement (SeaNet)}. Medical images always contain some irrelevant information that can disrupt the decision-making. For this reason, to solve ambiguous classification problems, it is essential to refine the features extracted by the CNN model. To this end, an attention module $G_A(F; \delta^s)$ has been proposed to mimic the clinical behavior by focusing on the key features for the prediction,  $G_A:F \rightarrow A$. In this case, the input of the attention module corresponds to the output feature map generated by the feature extractor, $F\in \mathbb{R}^{H \times W \times C}$. The proposed attention module works as a kind of autoencoder composed of $1\times1$ convolutions in which the filters are decreased and increased, respectively. Therefore, the feature maps obtained at the output of each of these convolution layers will have the same 
spatial dimension as the previous feature map, with the difference that the number of channels will have been changed to accomplish a combination of the features. In order to explore the dependencies existing among the different feature channels as well as the contextual information, the blocks called `Squeeze-and-Excitation' (SE) \cite{hu2018squeeze} were implemented between the different convolutional reduction layers of the attention module, see Figure \ref{SE}. 

The input to the SE block, $G \in \mathbb{R}^{H \times W \times R}$, is embedded into a $s \in \mathbb{R}^{1 \times 1 \times R}$ vector by a global average pooling (GAP) layer, which provides a global distribution of responses by channels. Note that the number of filters $R$, corresponds to the number of channels at the output of the convolutional layers of the attention module. In the following step, $s$ is transformed into $\hat{s}= \phi(W_2(\partial(W_1s)))$ where $\phi$ is the sigmoid activation function, $W_1 \in\mathbb{R}^{\frac{R}{r}\times R}$ and $W_2 \in\mathbb{R}^{R\times}\frac{R}{r}$ are the weights of two completely fully-connected layers (FC) and $\partial$ is the Relu activation function. The parameter $r$ is the reduction ratio for dimensionality reduction, in this case $r=4$, indicating the bottleneck. After the sigmoid activation, the activations of $\hat{s}$ are ranged to [0,1] and it is used to recalibrate the input $\textit{\textbf{G}}=[g_1,g_2,...,g_c]$ where $g_i \in \mathbb{R}^{HxW}$. The output feature map of this block is $\textit{\textbf{G}}_{se}=[\hat{s_1}g_1,\hat{s_2}g_2,...,\hat{s_c}g_c]$.

The last reduction layer of the attention module has the sigmoid as activation function to recalibrate the inputs and force the network to learn useful properties from the input representations. After increasing the number of filters to the same number as the input layer to this module, the output of the attention module is pondered with the output of the feature extractor obtaining a refined feature map $A \in \mathbb{R} ^{H \times W \times C}$.

\begin{enumerate} [B.]
\item \textit{Projection head module}
\end{enumerate}
In this paper, we instantiate a projection head network, $G_h : A \rightarrow Z$, that maps the representations \textit{A} to an embedding vector \textit{Z} where the classification stage is addressed in a lower-dimensional space. In this case, different configurations already applied in the literature have been tested in Section 5. In contrast to other widely used approaches such as the flattening of the activation volume resulting from the final convolutional block and the class prediction through consecutive fully-connected layers, the global max pooling (GMP) and the global average pooling (GAP) layers reduce the number of parameters decreasing the complexity of the model. At the end of the convolutional network, a softmax-activated dense layer
is applied to address the tumor region identification.

\subsection{Target model: WSI prediction}
\label{feature_extractor}

The target model aims to classify spitzoid lesions under an embedded-space paradigm using the biopsy-level labels for learning. To that end, our main goal is to find a compact embedding for the instances of a bag/WSI and combine these instance embeddings to a single embedding that represents the entire bag, see Figure \ref{target_model}.

Specifically, we denote each individual bag as $X^t_n=\left\lbrace x^t_{n,1},..., x^t_{n,i} ,x^t_{n,I_n} \right\rbrace$, where $x^t_{n,i}$ is the i-th predicted tumor instance by the source model and $I_n$ denotes the total number of predicted tumor region patches in a slide. Note that $I_n$ can vary across bags. Hence, the objective of the target model becomes to obtain the label of a slide ($\hat{Y}^t_n$) from the tumor instances predicted by the source model  ($x^t_{n,i}$), which can be defined as follows:

\begin{equation}
\centering
\hat{Y}^t_n = f( \left\lbrace x^t_{n,1},..., x^t_{n,i} ,...,x^t_{n,I_n}  \right\rbrace, 	\omega^t)  
\end{equation}where $\omega^t$ denotes the target model weights.

In order to find an embedding representation of each bag, we use the pre-trained backbone and the projection head module of the source model. In this manner, following an inductive learning strategy, the backbone already has prior knowledge concerning basic features of the histological database. After embedding each bag, $\textbf{h}_n$ = $G_h(G_A(G_f(X^t_n)))$, we obtain a C-dimensional feature vector for each instance. The bag label predictor $G_y : \left\lbrace \textbf{h}_{i} \right\rbrace _{i\in I_n} \rightarrow \hat{Y}^t_n $ aggregates the C-dimensional feature vectors $\left\lbrace \textbf{h}_{i} \right\rbrace _{i\in I_n}$ into a feature vector $Z_n \in \mathbb{R}^{1 \times C}$ representative of the bag. In the literature, there exist different aggregation functions such as batch global max pooling (BGMP) or batch global average pooling (BGAP). However, such functions are not flexible since they do not have trainable parameters. For this reason, in this work we use a trainable aggregation function \cite{ilse2018attention}. In this case, $G_y(\cdot ; \omega^t)$ is characterized by a set of trainable parameters $\textbf{V} \in \mathbb{R}^{L \times C}$ and $\textbf{w} \in \mathbb{R}^{L \times 1}$. The embedded feature vector per bag is obtained as $Z_n=\sum_{i \in I_n} a_i \cdot \textbf{h}_i$, where $a_i$ is defined as:

\begin{equation}
\centering
a_i=\frac{exp ( \textbf{w}^T tanh(\textbf{V}\textbf{h}_i) )}{\sum_{j \in I_n} exp (\textbf{w}^T tanh (\textbf{V}\textbf{h}_j)) }
\end{equation}

The attention-based aggregation function is differential and can be trained in a end-to-end manner using gradient descent. Additionally, the attention module not only provides a more flexible way to incorporate information from instances, but also enables us to localize informative tiles. The superiority of this aggregation function for spitzoid prediction will be shown in Section 5. Finally, the $Z_n$ vector attaches to the dense layer with a sigmoid function-activated neuron to obtain the prediction at the biopsy level.

\section{Ablation Experiments}
\label{seccion_experimental}

In this section, we present the results of the different experiments carried out to show the performance of the proposed approach for the different classification tasks: patch-level classification (source model) and WSI prediction (target model). Note that a comparison with the current state-of-the-art methods was not possible as there are no algorithms focused on histological images of spitzoid tumors. Additionally, no public databases of histological images with melanocytic neoplasms have been found to apply our algorithms.

\subsection{Database partitioning}

Making use of the spitzoid database (CLARIFYv1), we carried out a patient-level data partitioning procedure to separate training and testing sets, aiming at avoiding overestimating the performance of the system and ensuring its ability to generalize. Specifically, 30\% of patients were used to test the models, whereas the remainder of the database was employed to train the algorithm. To train the proposed models and optimize the hyperparameters involved in this process, the training set was divided following a cross-validation strategy. We used four validation cohorts to optimize both the source and the target models. To encourage the source model to select the most relevant tiles, we used an instance dropout over the non-tumor region, since these represent the majority class. Specifically, instances were randomly dropped during the training, while all instances were used during the model evaluation.

\subsection{Source model selection}

\begin{enumerate} [A.]
\item \textit{Backbone optimization}
\end{enumerate}

\begin{figure*}[hbt]
    \centering
    \includegraphics[width=13cm,height= 7.5cm]{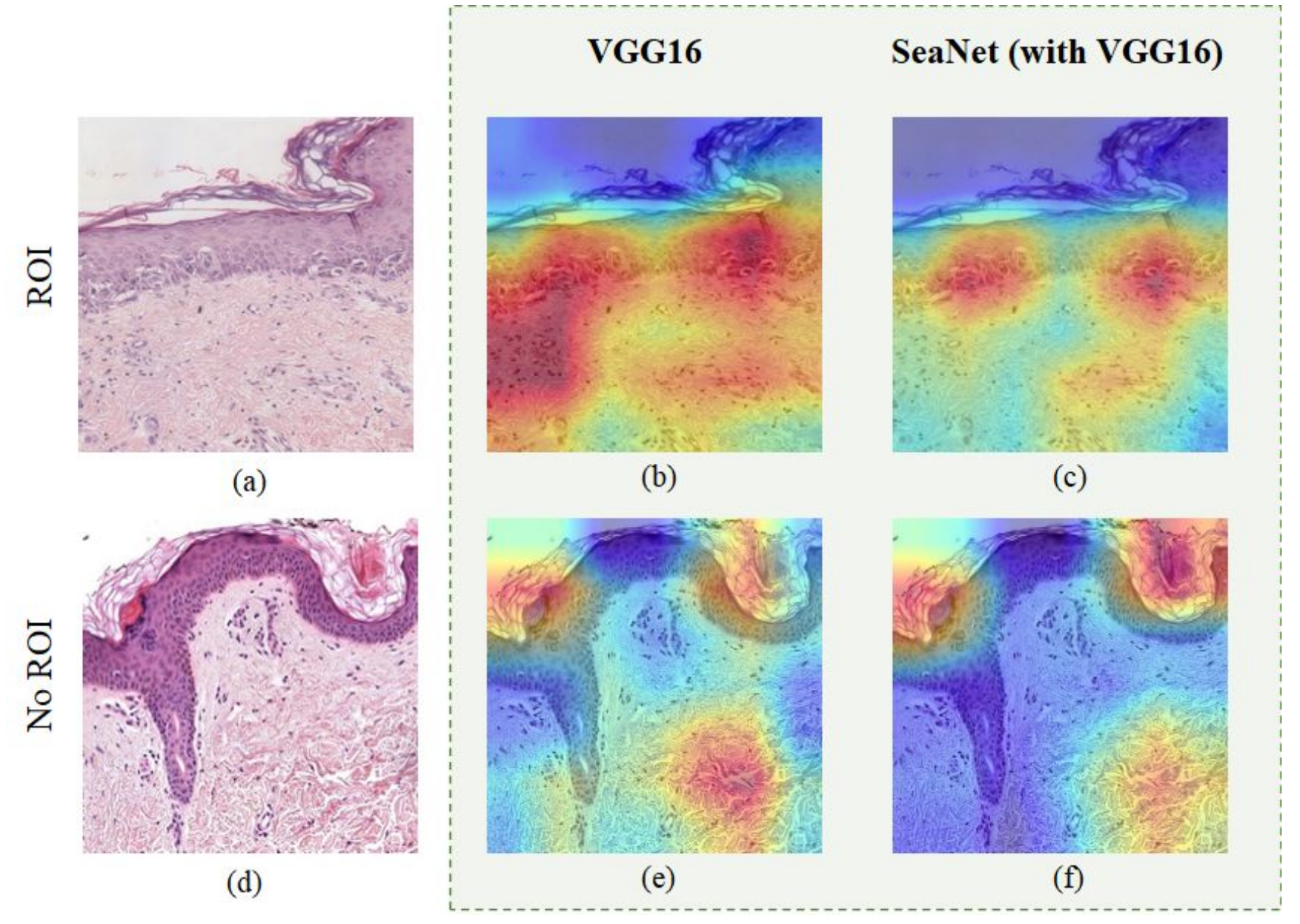}
    \caption{Original images (first column) and Class Activation Maps (CAMs) obtained using the VGG16 model (second column) and the SeaNet with VGG16 (third row) in two images correctly classified. ROI region is shown in the first row and no ROI in the second one.}
    \label{CAM_corect}
\end{figure*}

\begin{figure*}[htb]
    \centering
    \includegraphics[width=14cm,height= 8cm]{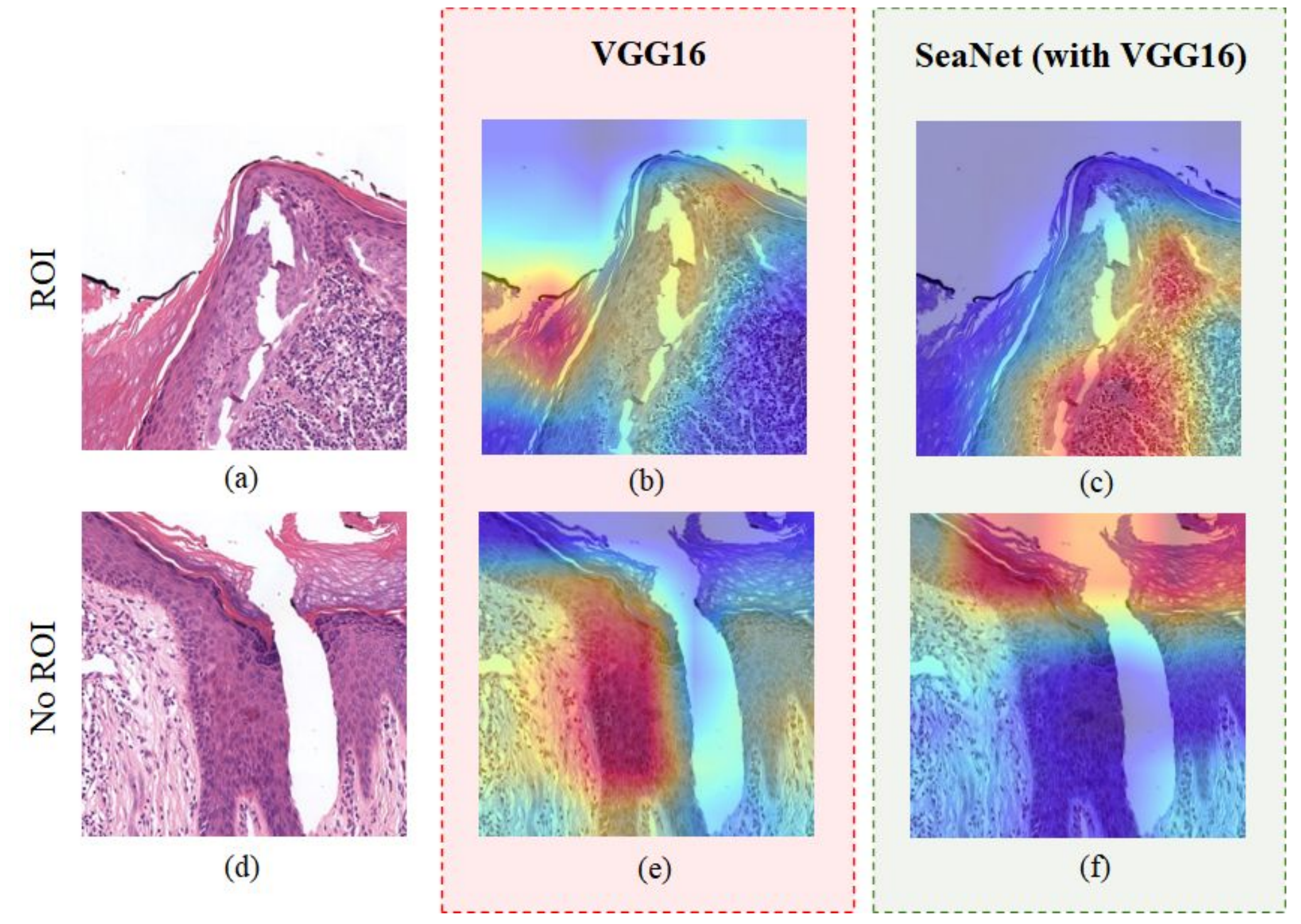}
    \caption{Original images (first column) and Class Activation Maps (CAMs) obtained with the VGG16 model (second column) and the SeaNet with VGG16 (third column). (b) and (e): Patches misclassified by the VGG16 model predicted as no ROI and ROI, respectively; (c) and (f): Patches well classified, ROI and No ROI respectively. }
    \label{CAM_incorect}
\end{figure*}

According to the literature for histopathological image analysis, we compared as feature extractors the well-known ResNet and VGG architectures since they have reported the best performance \cite{Hekler2019,wang2020automated}. Additionally, we applied the proposed feature refinement SeaNet, Squeeze and Excitation Attention Network, on each of these feature extractors in order to evaluate the enhancement introduced. To address an objective comparison of the proposed backbones, we kept constant the projection head module using a GAP layer. In Table \ref{backbone_prediction}, we contrast the validation results achieved by the different backbones trained in a binary-class scenario. The comparison was handled by means of different figures of merit, such as sensitivity (SN), specificity (SPC), positive predictive value (PPV), negative predictive value (NPV), F1-score (F1S), accuracy (ACC) and area under the ROC Curve (AUC). Note that the figures of merit listed above report the results for the average of the validation cohorts in the cross-validation process. Additionally, class activation maps (CAMs) were computed to highlight the regions of interest at patch-level in which the proposed source model paid attention to predict the samples, see Figure \ref{CAM_corect} and Figure \ref{CAM_incorect}. The backbone reporting the best performance during the validation stage was selected as the base encoder network to address the head projection optimization.

\begin{table*}[h]
\caption{Classification results reached during the validation stage with the proposed fine-tuned architectures. }
\label{backbone_prediction}
\setlength\tabcolsep{5 pt}
\small
\begin{center}
\begin{tabular}{ccccc}
\hline
\multicolumn{1}{l}{}{} & \textbf{VGG16} &  \textbf{SeaNet (with VGG16)}    & \textbf{RESNET}  & \textbf{SeaNet (with RESNET)} \\
\hline
\textbf{SN}   &  $0.8057 \pm 0.1247$    & $\textbf{0.8310} \pm \textbf{0.1061}$    & $0.8200 \pm 0.1223$  &  $0.7494 \pm 0.1736$ \\
\textbf{SPC}   & $0.9070 \pm 0.0343$    & $\textbf{0.9298} \pm \textbf{0.0185}$  &  $0.8850 \pm 0.0243$  &   $0.9290 \pm 0.0422$   \\
\textbf{PPV}   &  $0.8448 \pm 0.0856$     & $ \textbf{0.8814} \pm \textbf{0.0495}$ & $ 0.8061 \pm 0.1005$   & $ 0.8800 \pm 0.0316$  \\
\textbf{NPV} &   $0.8894 \pm  0.0649$     & $\textbf{0.9100} \pm  \textbf{0.0232}$      &  $0.8830 \pm  0.0761$   &  $0.8693 \pm  0.0516$    \\
\textbf{F1S}     & $0.8183 \pm 0.0865$    & $\textbf{0.8654} \pm \textbf{0.0805}$  &  $0.8022 \pm 0.1126$  &  $0.8100 \pm 0.0927$     \\
\textbf{ACC}    &   $0.8752 \pm 0.0357$  & $ \textbf{0.9031} \pm \textbf{0.0262}$  & $0.8611\pm 0.0558$   & $0.8770 \pm 0.0329$   \\
\textbf{AUC}  & $0.8600 \pm 0.0584$ & $\textbf{0.8810} \pm \textbf{0.0566}$    &   $0.8400 \pm  0.0813$      & $0.8500 \pm  0.0737$  \\
\hline
\end{tabular}
\end{center}
\end{table*}

\textbf{Training details.} All the contrasting approaches were implemented using Tensorflow 2.3.1 with Python 3.6. Experiments were conducted on the NVIDIA DGX A100 system. NVIDIA DGX A100 is the universal system for all artificial intelligence (AI) workloads, offering unprecedented compute density, performance, and flexibility in a 5 petaFLOPS AI system. All models were trained for 120 epochs using a learning rate of 0.001 with a batch size of 64. Stochastic gradient descent (SGD) optimizer was applied trying to minimize the binary cross-entropy (BCE) loss function at each epoch. The base model of the fine-tuned feature extractor was also optimized, selected to freeze the first convolutional block for VGG16 and setting all layers as trainable for ResNet.

\begin{enumerate} [ B.]
\item \textit{Head projection optimization}
\end{enumerate}
In this section, we report the validation performance using different projection head modules. Specifically, we compare a small multi-layer perceptron (MLP) with one hidden layer of 128 neurons non-linearly activated by the ReLU function, a global max-pooling (GMP) layer and a global average-pooling (GAP) layer, see Table \ref{valRes_head_module}. It is important to note that the comparison was conducted using the proposed SeaNet (with VGG16) backbone for all the scenarios. 

\begin{table*}[htb]
\caption{Classification results reached during the validation stage using different projection head modules.}
\label{valRes_head_module}
\setlength\tabcolsep{5 pt}
\small
\begin{center}
\begin{tabular}{cccc}
\hline
\multicolumn{1}{l}{}{} & \textbf{SeaNet+MLP}       & \textbf{SeaNet+GMP}  & \textbf{SeaNet+GAP} \\
\hline
\textbf{SN}         & $0.8716 \pm 0.3000$  & $\textbf{0.8729} \pm \textbf{0.0371}$   & $0.8310 \pm 0.1061$ \\
\textbf{SPC}        & $0.9076 \pm 0.0478$   & $0.9143 \pm 0.0131$ &  $\textbf{0.9298}  \pm  \textbf{0.0185}$\\
\textbf{PPV}        & $0.8460 \pm 0.1018$ &       $0.8589 \pm 0.0710$  & $\textbf{0.8814 } \pm  \textbf{0.0495}$  \\
\textbf{NPV}        & $0.9100  \pm 0.0348$ &    $\textbf{0.9140} \pm  \textbf{0.0283}$   & $0.9100  \pm  0.0232$ \\
\textbf{F1S}         & $0.8606  \pm 0.0655$ &        $\textbf{0.8708} \pm \textbf{0.0541}$  & $0.8654  \pm  0.0805$\\
\textbf{ACC}        & $0.8940 \pm 0.0320 $  &     $0.9020 \pm 0.0164$  & $ \textbf{0.9031}  \pm  \textbf{0.0262}$ \\
\textbf{AUC}        &  $0.8800 \pm 0.0391 $  &       $\textbf{0.8935} \pm \textbf{0.2490}$  &  $0.8810  \pm  0.0566$\\
\hline
\end{tabular}
\end{center}
\end{table*}
\textbf{Training details.} The same hardware and software systems as for the backbone section were used to optimize the head projection. Additionally, we use the same learning rate, batch size, loss function and number of epochs as in the previous section. In this case, we only changed the head projection. 

\subsection{Target model selection}
\begin{enumerate} [ A.]
\item \textit{WSI label predictor optimization}
\end{enumerate}
As mentioned throughout the manuscript, the backbone and the projection head module of the target model were optimized during the ROI selection, via the source model. After obtaining an embedded feature vector of each tile in a bag, it is necessary to implement an aggregation function. In this section, we compare the results, when three different aggregation functions were used: batch global max-pooling (BGMP), batch global average pooling (BGAP) and batch global attention summary (BGAS), Table \ref{valRes_target_model}.

\begin{table*}[htb]
\caption{Classification results reached during the validation stage using different aggregation functions: batch global max-pooling (BGMP), batch global average-pooling (BGAP) and batch global attention summary (BGAS).}
\label{valRes_target_model}
\setlength\tabcolsep{5 pt}
\small
\begin{center}
\begin{tabular}{cccc}
\hline
\multicolumn{1}{l}{}{} & \textbf{BGMP}       & \textbf{BGAP}  & \textbf{BGAS} \\
\hline
\textbf{SN}         & $0.5000 \pm 0.3953$  & $0.5833 \pm 0.3062$   & $\textbf{0.7500} \pm \textbf{0.2764}$ \\
\textbf{SPC}        & $ \textbf{0.9000} \pm \textbf{0.3953}$   & $0.8500 \pm 0.1658$ &  $0.8500  \pm  0.2764$\\
\textbf{PPV}        & $0.6250 \pm 0.4330$ &       $0.8375 \pm 0.1709$  & $\textbf{0.8667} \pm  \textbf{0.1414}$  \\
\textbf{NPV}        & $ 0.7625   \pm  0.1546$ &    $0.7848 \pm  0.1388$   & $\textbf{0.8869}  \pm  \textbf{0.1207}$ \\
\textbf{F1S}         & $0.5018  \pm 0.3873$ &        $0.6000 \pm 0.1541$  & $\textbf{0.7472}  \pm  \textbf{0.1473}$\\
\textbf{ACC}        & $0.7361  \pm 0.0977 $  &     $0.7361 \pm 0.0417$  & $ \textbf{0.8229}  \pm  \textbf{0.0262}$ \\
\textbf{AUC}        &  $0.7000 \pm 0.1744 $  &       $0.7167 \pm 0.0841$  &  $\textbf{0.8000}  \pm \textbf{0.0963}$\\
\hline
\end{tabular}
\end{center}
\end{table*}

\textbf{Training details.} In order to generate bags and train the algorithms, a maximum of 300 image patches were randomly extracted from the source model prediction. The whole models were re-trained during 100 epochs using a learning rate of 0.001 and a batch size of 1 slide. Stochastic gradient descent (SGD) optimizer was applied trying to minimize the binary cross-entropy (BCE) loss function at every epoch.

\section{Prediction Results} \label{prediction}
In this section, we show the quantitative and qualitative results achieved by the proposed strategies during the prediction of the test set. For both methods developed in this work, ROI selection and WSI classification, predictions were performed using the architectures with the best performance during the validation stage.

\textbf{Quantitative results}. Table \ref{test_prediction} shows the results reached in the test prediction for the proposed source and target models.
\begin{table}[htb]
\caption{Classification results reached during the prediction stage. The proposed source model (SM) was composed of the SeaNet (with VGG16) + GMP. The proposed target model (TM) used the BGAS layer as an aggregation function.}
\label{test_prediction}
\setlength\tabcolsep{5 pt}
\small
\begin{center}
\begin{tabular}{ccc}
\hline
\multicolumn{1}{l}{}{} & \textbf{SM} & \textbf{TM}      \\
\hline
\textbf{SN}         & 0.9285  & 0.6700 \\
\textbf{SPC}        &0.9202  &0.8900\\
\textbf{PPV}        &  0.8622  &  0.8000  \\
\textbf{NPV}        &  0.9599    &  0.8000    \\
\textbf{F1S}         & 0.8942   & 0.7300  \\
\textbf{ACC}        & 0.9231  & 0.8000 \\
\textbf{AUC}        & 0.9244  & 0.7800 \\
\hline
\end{tabular}
\end{center}
\end{table}

\begin{figure*}[h]
    \centering
    \includegraphics[width=12cm,height= 14cm]{ 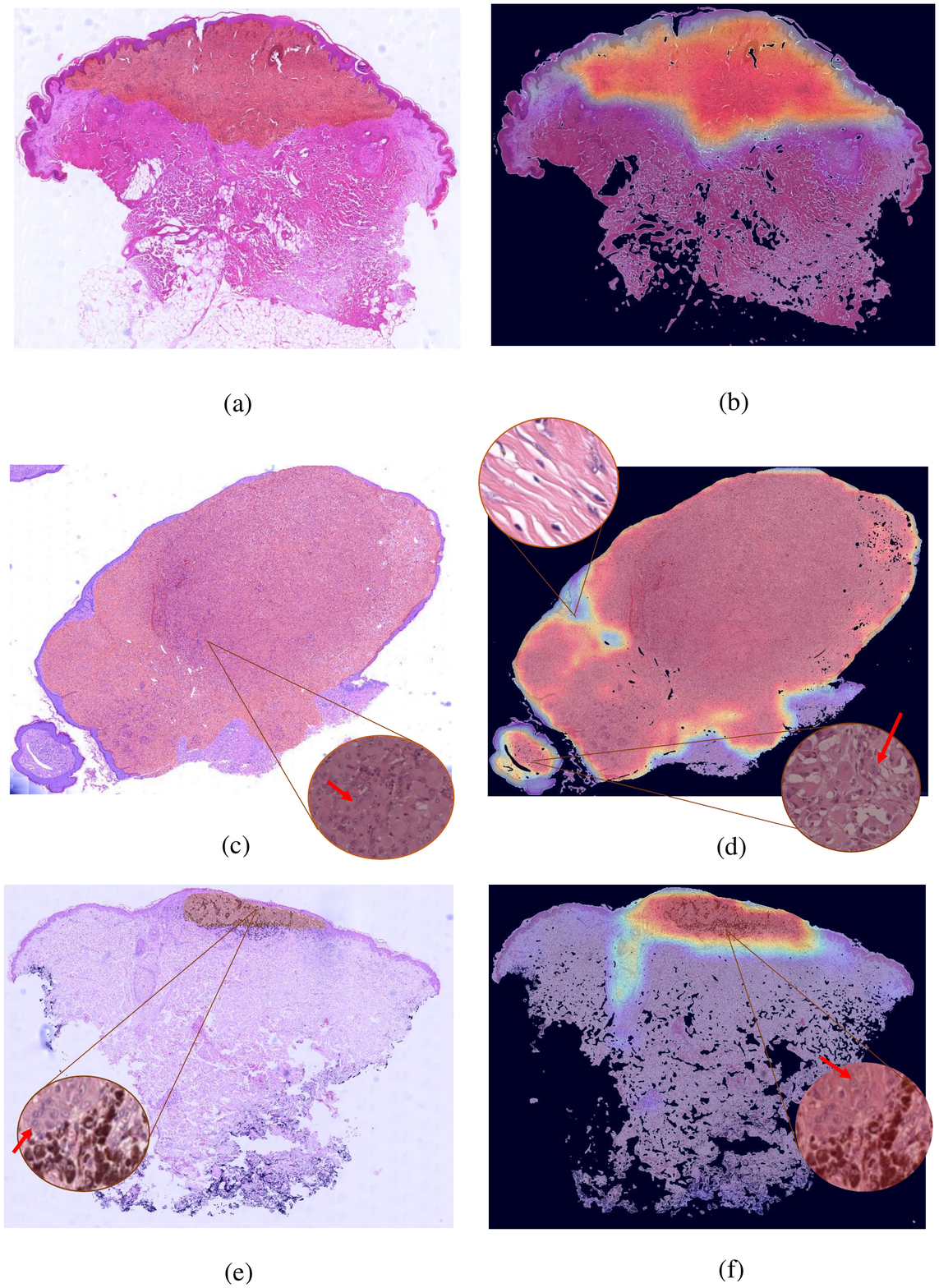}
    \caption{Whole slide image level prediction for the source model (ROI estimation). (a), (c), (d) Manual annotations, (b), (d), (f) System predictions.}
    \label{fig:probability maps}
\end{figure*}

\textbf{Qualitative results}. To qualitatively show the performance of the ROI selection model, we obtained probability heatmaps of representative samples indicating the presence of tumor region in the WSIs, Figure \ref{fig:probability maps}.

\begin{figure*}[h]
    \centering
    \includegraphics[width=14cm]{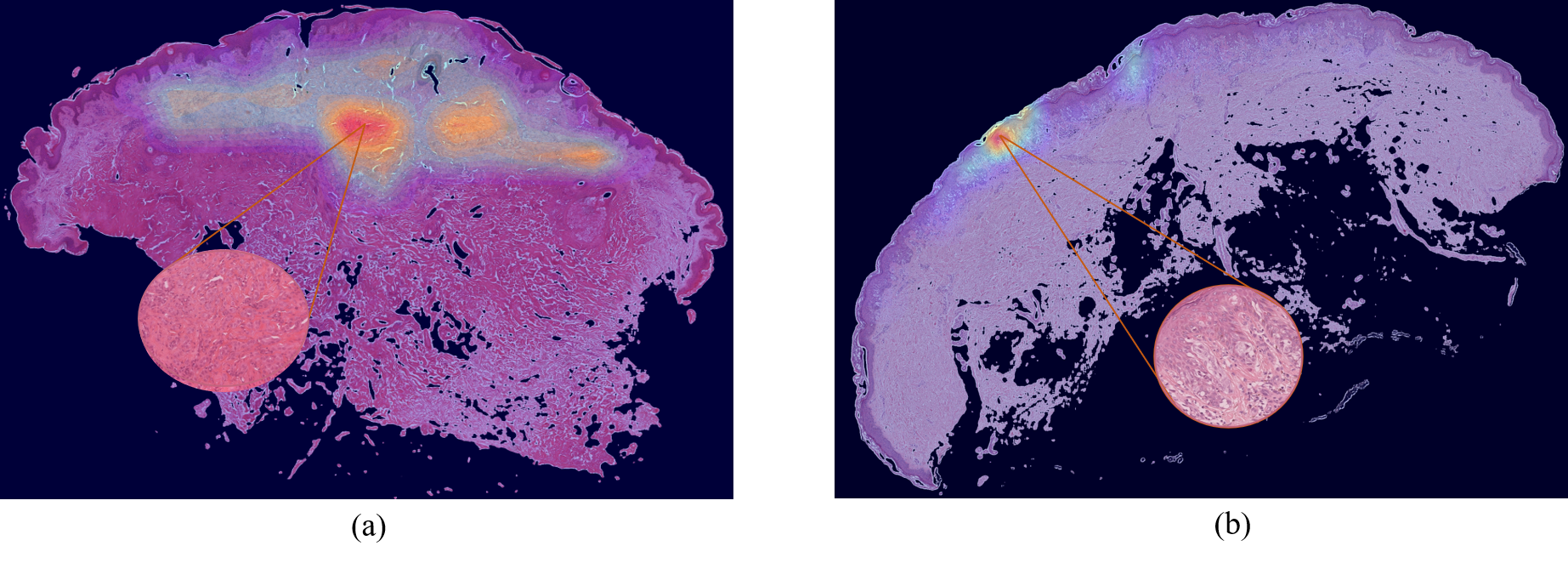}
    \caption{Visualization of the attention weights of the bag aggregation function in heat maps. (a) Benign sample; (b) Malignant sample.}
    \label{fig:attention_weights}
\end{figure*}

In the probability maps, for each pixel, the predicted probabilities for the ROI are estimated by bilinearly interpolating the predicted probabilities of the closest patches in terms of euclidean distance to the center of the patches. In addition, using these heatmaps, we visualize the distribution of attention weights, which were calculated for correctly classified cases into benign and malignant neoplasms, see Figure  \ref{fig:attention_weights}.

\section{Discussion} \label{seccion_experimental}
In this section, we make reference to the main contributions detailed throughout the paper and we review the obtained results. 

In contrast to the state-of-the-art studies for histological images classification, in which the input of the prediction model is the tumoral region annotated by the pathologist, in this paper, we propose a framework able to first automatically select neoplastic regions of interest and then predict the malignancy or benignity of spitzoid neoplasms. Note that no previous studies seem to have been proposed any automated method for the detection of these challenging neoplasms.

\subsection{Source model: ROI selection}
\begin{enumerate} [ A.]
\item \textit{About the ablation experiment}
\end {enumerate}

\textbf{Backbone selection}. As a first stage, we carried out an optimization of the feature extractor for the selection of the tumor regions. Considering the limited amount of available samples, we decided to use the fine-tuning technique on the VGG16 and RESNET architectures. Particularly, from Table \ref{backbone_prediction} we can observe that the use of sequential approaches (VGG16) provided slightly better results than architectures with residual blocks (RESNET). This fact is evidenced in several works of the literature for histopathological analysis where the sequential models used to outperform residual ones \cite{Silva-Rodriguez2020}. Additionally, the proposed SeaNet module, characterized by the refinement of the features via convolutional attention blocks, reported a significant outperforming. Specifically, the SeaNet module via fine-tuning VGG16 architecture achieved the best results. The use of the attention module provides more distinctive feature maps and allows a considerable reduction in the incidence of false positive and false negative samples, leading to improve global metrics. Aiming at qualitatively observing the enhancements introduced by the refinement module, the CAMs of the best models (SeaNet with VGG16 and VGG16 alone) were obtained for correctly classified images (see Figure \ref{CAM_corect}) and for images misclassified by the VGG16 model (see Figure \ref{CAM_incorect}). In Figure \ref{CAM_corect}, we can see that both for the prediction of patches belonging to the tumor region (a) and for non-tumoral ones (b), the SeaNet activations are focused on smaller regions. For the ROI prediction, the SeaNet (with VGG16) model is mainly focused on the pagetoid pattern present within the epidermis, defining the region as tumor. However, the VGG16 model extends its activations to lymphocytes found within the dermis. In this case, the lymphocytes do not necessarily determine that the region is tumorous, since this small amount of lymphocytes can also be found in healthy regions. Therefore, the VGG16 model without the attention module introduces certain noise in the prediction. Regarding the prediction of non-tumor regions, both models are focused on the epidermis and stromal region of the dermis.
Regarding the cases where VGG16 misclassifies tumor regions, Figure \ref{CAM_incorect} (b), the activations are focused on the epidermis region. In this case, the epidermis region has no patterns indicative of a melanocytic lesion, but for a correct classification, the activations would have to be focused on the melanocyte aggregate found in the upper region, as in the case of the SeaNet model, see Figure \ref{CAM_incorect} (c). In this region, we find a large number of melanocytic cells with a high concentration of lymphocytes indicating an inflammatory reaction to a tumor region. For the case of the non-tumor region shown in Figure \ref{CAM_incorect} (d), the VGG16 model erroneously predicts it by focusing on the melanocytic cells found in the epidermis, see Figure \ref{CAM_incorect} (e). Normally, in healthy skin, the dermo-epidermal junction is composed of isolated melanocytic cells with a certain spacing between them. It is representative of a tumor when these cells ascend to the upper layers of the epidermis forming what is known as a pagetoid pattern or infiltrate the dermis forming nests. Furthermore, in this case, the epidermis has no patterns that would be representative of a melanocytic lesion. Unlike the VGG model, the SeaNet (with VGG16) model reports its activations in the epidermal region and based on it establishes the correct prediction, classifying this patch as non-characteristic of a spitzoid lesion, see Figure \ref{CAM_incorect} (f).

In any case, the inclusion of the proposed attention module outperforms the popular pre-trained architectures of the state of the art and reduces the number of noisy patches used as input to the target model.

\textbf{Projection head module selection}. After optimizing the backbone, we proceeded to select the projection head module that provided the best results. For this purpose, we tested three projection head modules: multilayer perceptron (MLP), global average-pooling (GAP) and global max-pooling (GMP). Table \ref{valRes_head_module} shows that the modules based on GAP and GMP provide very similar and significantly better results than those reported by the MLP. The outperforming of GMP and GAP compared to the fully-connected configuration could be explained by the reduction in the number of weights to be optimized, making the model simpler and more capable of generalizing to new images. Comparing the results provided by GAP and GMP, we can conclude that they are very similar. The main difference between these techniques lies in the way of squeezing the spatial dimension. While GMP considers only the maximum value for the feature map, in the GAP layer the whole spatial region contributes to its output. This explains why the GMP layer enhances SN results and the GAP layer improves SPC results. With the GMP layer, it is more likely to well classify a patch belonging to the tumor region, even if it contains a minimal tumor region. However, GAP takes into account the whole context so that regions with small tumor areas are likely to be discarded. Although both show a very similar result, global metrics such as F1S and AUC exhibit a slight improvement with the GMP layer. Therefore, the GMP layer will be preferred as the optimal head projection module.
\begin{enumerate} [ B.]
\item \textit{About the prediction results}
\end{enumerate}
Table \ref{test_prediction} shows the results reached by the proposed ROI selection model. All the metrics reported here outperform those obtained in the validation phase. Figure \ref{fig:probability maps} shows the probability maps for the lesion region of three test samples. The majority of the lesion regions predicted by the algorithm are depicted in Figure \ref{fig:probability maps} (b), in which the prediction is completely in line with the annotation performed by the pathologists, Figure \ref{fig:probability maps} (a). Some activation maps, such as those shown in the Figure \ref{fig:probability maps} (d), predict certain areas annotated by the pathologists are non-tumor regions. However, if we visualize the expansion of the areas where there are no activations, we can see that there are no melanocytic nests characteristic of the lesion, and therefore, we may be facing a discontinuity of the lesion as explained in Section \ref{material}. In contrast, in the lower part of Figure \ref{fig:probability maps} (d), there are activations of tumor regions that have not been annotated by expert pathologists, see Figure \ref{fig:probability maps} (c). However, if these regions are enlarged, it can be concluded that tumor cells are present. At times, due to the large amount of material in a lesion, pathologists can overlook some tumor areas. In the case of Figure \ref{fig:probability maps} (e) and (f), there is also some discrepancy between the annotations performed by the pathologists and the activations predicted by the model. In these figures, we find melanocytic cells with melanosomes that give them their characteristic brown color. It is difficult to differentiate these tumor cells from melanophages (cells with brown staining and all of the same size) that are not tumor cells. In this case, if we zoom the activations of the algorithm (Figure \ref{fig:probability maps} (f)) in those regions not annotated by the pathologist, we can see that there are also tumor cells. Therefore, the developed algorithm could help the decision-making in cases where there is ambiguity for the pathologists. In this context, the developed method allows to enhance the detection of tumor areas.

\subsection{Target model: WSI prediction}
\begin{enumerate} [ A.]
\item \textit{About the ablation experiment}
\end {enumerate}
\textbf{WSI label predictor optimization}. As discussed throughout the document, the backbone used by the target model was optimized during the selection of the source model. Therefore, in this case, it was only necessary to optimize the aggregation function required to perform a prediction using a MIL approach. From Table \ref{valRes_target_model}, we can observe that the use of the feature average of all patches containing a bag to obtain the embedded representation provides the best results (BGAP and BGAS aggregation functions). Additionally, the BGAS aggregation function improves the results provided by BGAP thanks to the introduction of optimized attention weights by updating the bag-level predictor weights ($\omega^t$), achieving a validation accuracy of 0.8229. Therefore, we can conclude that the introduction of the attention module allows focusing on more relevant patterns, thus improving the final classification.

\begin{enumerate} [ B.]
\item \textit{About the prediction results}
\end{enumerate}

Table \ref{test_prediction} shows the results reached by the proposed target model in the test set. The results are in line with those obtained in the validation phase. Figure \ref{fig:attention_weights} shows the attention weights of the BGAS aggregation function for benign (Figure \ref{fig:attention_weights} (a)) and malignant (Figure \ref{fig:attention_weights} (b)) samples. The attention weights were normalized between 0 to 1 in each bag. The red regions in the attention weight maps represent the highest contribution for classification in each bag. Therefore, the bag class label is predicted by only using instances for which the attention values are large. In the case of a benign sample (Figure \ref{fig:attention_weights} (a)), the regions contributing to the class establishment are distributed over a wide area of the lesion, being these areas aggregates of melanocytes. However, the large attention weights for a malignant lesion are focused on small region characteristics of malignancy (in this case pagetoid pattern) as shown in Figure \ref{fig:attention_weights} (b).

\section{Conclusion}

In this work, we propose an inductive transfer learning framework able to perform both ROI selection and malignant prediction in spitzoid melanocytic lesions using WSIs. Our proposed framework is composed of a source model in charge of selecting the patches with characteristic lesion patterns. The source model introduces an attention module able to refine the features of the latent space to maximize the classification agreement. Using the backbone of the source model as a patch-level feature extractor and under a multiple instance learning approach, the target model predicts the malignancy degree by taking as input the tumor patches predicted by the first model. This innovative approach carried out in an end-to-end manner has reported promising results for both ROI selection and WSI classification, achieving a testing accuracy of 0.9231 and 0.8000 for the source and the target models, despite the limited number of samples. In this way, our framework bridges the gap with respect to the development of automatic diagnostic systems for spitzoid melanocytic lesions. In future research lines, the efforts should focus on improving the discrimination of the malignancy and benignity with the acquisition of new samples and enhancements to the implemented attention module in the multiple instance learning approach.

%-----------------------------  TABLES -----------------------------%
% \begin{table}[h]
% \centering
% \begin{tabular}{l l l}
% \hline
% \textbf{Treatments} & \textbf{Response 1} & \textbf{Response 2}\\
% \hline
% Treatment 1 & 0.0003262 & 0.562 \\
% Treatment 2 & 0.0015681 & 0.910 \\
% Treatment 3 & 0.0009271 & 0.296 \\
% \hline
% \end{tabular}
% \caption{Table caption}
% \end{table}

%-----------------------------  EQUATIONS -----------------------------%
% \begin{equation}
% \label{eq:emc}
% e = mc^2
% \end{equation}
\section*{Acknowledgements}

We gratefully acknowledge the support from the Generalitat Valenciana (GVA) with the donation of the DGX A100 used for this work, action co-financed by the European Union through the Operational Program of the European Regional Development Fund of the Comunitat Valenciana 2014-2020 (IDIFEDER/2020/030).

\section*{Funding}

This work has received funding from Horizon 2020, the European Union's Framework Programme for Research and Innovation, under grant agreement No.  860627 (CLARIFY), the Spanish Ministry of Economy and Competitiveness through projects PID2019-105142RB-C21 (AI4SKIN) and SICAP (DPI2016-77869-C2-1-R), Instituto de Salud Carlos III by the project PI20/00094 and GVA through project PROMETEO/2019/109. The work of Rocío del Amor and Laëtitia Launet has been supported by the Polytechnic University of Valencia (PAID-01-20) and  the European Union’s Horizon 2020 research and innovation programme under the Marie Skłodowska Curie grant agreement No 860627, respectively.

\section*{Conflict of interest}

The authors declare that they have no conflict of interest.



%% References



%% New version of the num-names style
\bibliographystyle{elsarticle-num}
\bibliography{refs.bib}






% \end{thebibliography}

\end{document}